# CRYSTALS AS A SOURCE OF GAMMA QUANTA

S.S. Georgadze

I. Javakhishvili Tbilisi State University, Tbilisi, Georgia

georgadzesusi@gmail.com

Conditions for the appearance of a nuclear exciton in a crystal consisting of excited and unexcited nuclei of a given type are determined. The total probabilities for spontaneous $\gamma$-emission by a single excited nucleus or by an arbitrary number of excited nuclei in the crystal are derived. It is shown that the formation of a nuclear exciton is connected with an increase of the width of the emitting level and with the concentration of the radiation in a narrow solid angle.

## INTRODUCTION

The main properties of spontaneous emission of excited nuclei are well known. The angular distribution of the $\gamma$ quanta radiated by a free unpolarized nucleus is isotropic. The lifetimes have been measured for a large number of low-lying exciting states of the nuclei. As a rule, the lifetime of nuclei situated in a medium differs little from the lifetime of free nuclei. Nonetheless, it seems to us that in principle there exists a possibility for the occurrence of an effect connected with the appreciable change in the lifetime of nuclei placed in a crystal. Such a possibility is realized only if certain necessary conditions are satisfied.

It is necessary first that the excited nuclei be placed in a crystal consisting of nuclei of the same sort in the ground state. In this case they quantum emitted by a single nucleus without recoil (Mossbauer effect) can be absorbed by another nucleus of the same sort. Therefore, in principle, the excitation becomes capable of propagating through the entire crystal (nuclear exciton). The wavelength of nuclear radiation in the energy range of interest to us $\lambdabar \ll a$ -crystal lattice constant. Therefore, the intensity of the radiation field in all the lattice sites that are closest to the radiating nucleus decreases like ($\lambdabar^2 / a$). It follows hence that although a nuclear exciton is possible in principle, this possibility is not realized in practice under usual conditions.

The situation is changed if all the nuclei are in phase, which, as will be shown below, corresponds to the condition $\mathbf{K} = 2\pi\mathbf{b}$, where $\mathbf{b}$-reciprocal lattice vector and $\mathbf{K}$-wave vector of the radiation. In this case the propagation of the radiation occurs in the direction of $\mathbf{b}$, i.e., it becomes one-dimensional. In one-dimensional propagation, the excitation intensity remains practically unattenuated over a distance ~ a, and consequently, creation of a nuclear exciton becomes possible. Of course, the nuclear exciton can exist only if the lifetime of the excited nucleus is not too small, namely $\tau > a/c \sim 10^{-18}$ sec. This criterion is satisfied in all known cases. These are similar to the macroscopic channeling in magnetic systems and in an intense electromagnetic wave which have been studied in [9–41] and [42-73].

In the present paper we study the conditions which must be satisfied for a nuclear exciton to exist. We obtain the $\gamma$-quantum radiation width for a crystal in which a nuclear exciton propagates.

## 1. RADIATION OF A NUCLEUS IN A CRYSTAL

We consider the condition for the occurrence of a nuclear exciton in the case of one excited nucleus placed in an ideal single crystal consisting of nuclei of the same sort. We assume that the Mossbauer conditions are satisfied. Calculation of the probability of the radiation of the nucleus in the crystal is best carried out by a method developed by Heitler [1]. Let $b_{\mathbf{K}\lambda}(t)$ be the amplitude of the probability of emission of a quantum with wave vector $\mathbf{K}$, frequency $\omega_{\mathbf{K}}$, and polarization $\lambda$, and let $b_1(\mathbf{r}_j, t)$ be the amplitude of the probability of finding an excited nucleus at the site j at the instant of time t. As follows from [1,2], these amplitudes are connected by the following equations ($\hbar = c = 1$)

$$i\dot{b}_{\mathbf{K}\lambda}(t) = H_{\mathbf{K}\lambda} e^{i(\omega_0 - \omega_{\mathbf{K}})t} \sum_j e^{i\mathbf{K}\mathbf{r}_j} b_1(\mathbf{r}_j, t),$$
$$i\dot{b}_1(t) = \sum_{\mathbf{K}\lambda} H_{\mathbf{K}\lambda}^* e^{-i(\omega_0 - \omega_{\mathbf{K}})t} b_{\mathbf{K}\lambda} e^{-i\mathbf{K}\mathbf{r}_i}; \quad (1)$$
$$b_1(0,0) = 1, \quad b_1(\mathbf{r}_i, 0) = 0, \quad \mathbf{r}_i \neq 0; \quad b_{\mathbf{K}\lambda}(0) = 0,$$

where $H_{\mathbf{K}\lambda}^* = h_{\mathbf{K}\lambda} e^{-W/2}$; $h_{\mathbf{K}\lambda}$ are the matrix elements for the radiation of a free nucleus, $e^{-W/2}$ is the Moss bauer factor [3], and the $\mathbf{r}_i$ are measured from the location of the excited nucleus.

The case of practical interest is when Ka » 1. We shall henceforth consider a crystal with cubic symmetry. Generalization to the case of a crystal with arbitrary symmetry entails no difficulty.

It can be verified that for arbitrary $\mathbf{K}$ and for nuclei that are close to the excited nucleus the amplitude $b_1(\mathbf{r}_j, t)$ is small like $(Ka)^{-1} \ll 1$. Indeed, let us rewrite (1) in the form

$$i\dot{b}_{\mathbf{K}\lambda}(t) = H_{\mathbf{K}\lambda} e^{i(\omega_0 - \omega_{\mathbf{K}})t} \left[ b_1(0,t) + \sum_{j \neq 0} e^{i\mathbf{K}\mathbf{r}_j} b_1(\mathbf{r}_j, t) \right],$$
$$i\dot{b}_1(t) = \sum_{\mathbf{K}\lambda} H_{\mathbf{K}\lambda}^* e^{-i(\omega_0 - \omega_{\mathbf{K}})t} e^{-i\mathbf{K}\mathbf{r}_i} b_{\mathbf{K}\lambda}(t). \quad (1')$$

In the case of arbitrary K the system (1') can be solved by iteration. As a result we obtain

$$b_1(\mathbf{r}_j,t) = \frac{\gamma_R e^{-W/2}}{2\omega_0 |\mathbf{r}_i|} \exp\left\{-\frac{\Gamma_R}{2}(t-|\mathbf{r}_i|)\right\} \frac{2}{i\Gamma_R}, \quad t > |\mathbf{r}_i|,$$
$$b_1(\mathbf{r}_j,t) = 0, \quad t > |\mathbf{r}_i|.$$
(2)

From (2) we see that the total radiation width $\Gamma_R$ is of the form

$$\Gamma_R = \gamma_R (1 + O(Ka)^{-2}).$$
(3)

i.e., in the case of arbitrary $\mathbf{K}$ and Ka » 1 the quantity $\Gamma_R$ differs little from $\gamma_R$, the radiation width for the free nucleus.

The situation changes noticeably when $\mathbf{K} = 2\pi\mathbf{b}$, where $\mathbf{b}$ is the reciprocal lattice vector. In this case, as can be seen from (1), the amplitudes $b_1(\mathbf{r}_i,t)$ do not depend on $\mathbf{r}_i$. Let us write the second equation of (1) in the form

$$i\dot{b}_1(t) = \sum_{\mathbf{K}\lambda} H_{\mathbf{K}\lambda}{}^* e^{-i(\omega_0 - \omega_{\mathbf{K}})t} b_{\mathbf{K}\lambda}(t) \sum_i e^{-i\mathbf{K}\mathbf{r}_i};$$
$$b(t) = \sum_i b_1(\mathbf{r}_i,t), \quad b(0) = 1.$$
(4)

We seek $b = e^{-\Gamma_R t/2}$. Inasmuch as $b_1(\mathbf{r}_i,t)$ does not depend on $\mathbf{r}_i$ when the following condition is satisfied

$$\mathbf{K} = 2\pi\mathbf{b},$$
(5)

we get

$$\sum_j e^{i\mathbf{K}\mathbf{r}_j} b_1(\mathbf{r}_j,t) = b(t) \frac{1}{N} \sum_j e^{i\mathbf{K}\mathbf{r}_j}$$
(6)

Solving (1) with account of (4)-(6), we obtain

$$\Gamma_R = 2\pi \sum_\lambda \rho_\lambda(\omega_0) \int d\Omega_{\mathbf{K}} |H_{\mathbf{K}\lambda}|^2 \left|\sum_j e^{i\mathbf{K}\mathbf{r}_j}\right|^2 \frac{1}{N},$$
(7)

where $\rho_\lambda(\omega_0) d\Omega_{\mathbf{K}}$ is the density of the final states of the emitted $\gamma$ quantum with polarization $\lambda$. The quantity $\Gamma_R$ has the meaning of the width of the $\gamma$-quantum radiation of the entire cyrstal. If the nuclei are not polarized, then

$$\Gamma_R = \frac{\gamma_R}{4\pi} e^{-W} \frac{1}{N} \int d\Omega_{\mathbf{K}} \left|\sum_j e^{i\mathbf{K}\mathbf{r}_j}\right|^2.$$
(8)

If conversion takes place in addition to the emission of the $\gamma$ quanta, then the total lifetime of the nucleus in the crystal is $\tau = 1/\Gamma_{tot}$ where $\Gamma_{tot} = \gamma_C + \gamma_R$, and $\gamma_C$ is the conversion width.

As in the theory of x-ray diffraction, we can show that $\Gamma_R \sim N^{1/3}$ [4]. From (8) we see also that radiation in the direction of **K** has the following probability per unit time:

$$\frac{dw}{d\Omega_\mathbf{K}} = \frac{\gamma_R}{4\pi} e^{-W} \frac{1}{N} \left| \sum_j e^{i\mathbf{K}\mathbf{r}_j} \right|^2. \tag{9}$$

If condition (5) is satisfied exactly, expression (9) becomes

$$\frac{dw}{d\Omega} = \frac{\gamma_R}{4\pi} e^{-W} \frac{1}{N}. \tag{10}$$

Comparing (8) and (10) we see that the intensity of $\gamma$-quantum emission lies in a narrow solid angle $\Delta\theta \sim 1/N^{2/3}$ near the direction of **b**.

The probability of emission of the y quantum by the entire crystal is larger than the probability of emission by a free nucleus by a factor $\sim e^{-W} N^{1/3}$. This effect is connected with the collectivization of the nuclear excitation in the crystal when the resonance condition (5) is satisfied. In other words, the lifetime of an excited nucleus placed in a crystal consisting of like nuclei will decrease when the conditions necessary for the production of a nuclear exciton are satisfied. Of course, the decrease in the lifetime of the excited nucleus will be appreciable only in the case when $\Gamma_R \geq \gamma_C$.

Formally (10) contains the total number of nuclei in the crystal. In fact, however, owing to the extinction phenomenon, which is well known in x-ray physics [5], the number N of nuclei is limited. As a result, the effective number of nuclei $N_{eff}$ will be $\sim (l_{ext})^3 n$, where *lext* is the extinction length and n the density of nuclei in the crystal.

We have considered above the radiation from a single nucleus placed in a crystal. Of practical interest is the case when the number of nuclei is arbitrary.

## 2. RADIATION IN THE CASE OF AN ARBITRARY NUMBER OF EXCITED NUCLEI

Let the number of excited nuclei in the crystal be arbitrary. We take into consideration the fact that an arbitrary excited nucleus interacts with an unexcited nucleus because the excited nucleus emits a y quantum, while the unexcited nucleus absorbs the y quantum ($\gamma$-quantum exchange). Then the Hamiltonian of the system can be represented in the following form [6] (we neglect conversion)

$$H = H_0 + H_{int}$$
$$H_0 = \frac{\omega_0}{2} \sum_i (a^+_{i+1/2} a_{i+1/2} - a^+_{i-1/2} a_{i-1/2}),$$
$$H_{int} = -\frac{1}{2} \sum_{ij} \sum_{\mathbf{K}\lambda} |H_{\mathbf{K}\lambda}|^2 \left( \frac{1}{\omega_\mathbf{K} - \omega_0 - i\gamma/2} - \frac{1}{\omega_\mathbf{K} - \omega_0 + i\gamma/2} \right)$$
$$\times \left[ \exp(i\mathbf{K}(\mathbf{r}_i - \mathbf{r}_j)) a^+_{i+1/2} a^+_{j+1/2} a_{j-1/2} a_{i-1/2} + c.c. \right],$$
(11)

where $a^+_{j+1/2}$ is the operator for the creation of an excited nucleus in the site j, $a^+_{j-1/2}$ is the operator for the creation of an unexcited nucleus, and Hint the operator of interaction between the excited and the unexcited nuclei via y-quantum exchange.

Inasmuch as it is impossible to create excited and unexcited nuclei in a simultaneously single site, the operators $a^+_{j\nu} a_{j\nu} (\nu = \pm 1/2)$ should be of the Ferm1 type. The corresponding relations are

$$a_{i\nu} a^+_{j\nu_1} + a^+_{j\nu_1} a_{i\nu} = \delta_{ij} \delta_{\nu_1 \nu}$$
(12)

The Hamiltonian (11) can be rewritten, using (12), in the form

$$H = H_0 - \frac{1}{2} \sum_{ij} B(\mathbf{r}_i - \mathbf{r}_j) a^+_{i+1/2} a^+_{j-1/2} a_{j+1/2} a_{i-1/2};$$
$$B(\mathbf{r}_i - \mathbf{r}_j) = 2 \sum_{\mathbf{K}\lambda} |H_{\mathbf{K}\lambda}| e^{i\mathbf{K}(\mathbf{r}_i - \mathbf{r}_j)} \left( \frac{1}{\omega_\mathbf{K} - \omega_0 - i\gamma/2} - \frac{1}{\omega_\mathbf{K} - \omega_0 + i\gamma/2} \right).$$
(13)

We diagonalize the Hamiltonian (13) by means of the Bogolyubov canonical transformation [7]. We introduce the Fermi amplitudes $\alpha_i$ and $\beta_j$. The initial amplitudes are expressed in terms of $\alpha_i$ and $\beta_j$ with the aid of the linear relations

$$a_{j+1/2} = u\alpha_j + v\beta_j^+, \quad a_{j-1/2} = v\alpha_j - u\beta_j^+,$$
(14)

where u and v are coefficients satisfying the normalization condition $u^2 + v^2 = 1$. The operator $\widehat{N}$ of the total number of nuclei is expressed in terms of the new amplitudes with the aid of the relations

$$\widehat{N} = N + \sum_j a_j^+ a_j - \sum_j \beta_j^+ \beta_j. \tag{15}$$

It is also easy to verify that the operators $H_0$ and $H_{int}$ take the form

$$H_0 = \frac{\omega_0}{2}(v^2 - u^2)N - \frac{\omega_0}{2}(v^2 - u^2)\sum_j (a_j^+ a_j + \beta_j^+ \beta_j) + uv\omega_0 \sum_j (a_j^+ \beta_j^+ + \beta_j a_j),$$

$$H_{int} = -\frac{1}{2}\sum_{ij} B(\mathbf{r}_i - \mathbf{r}_j)[u^2 v^2 - 2u^2 v^2 (a_j^+ a_j + \beta_j^+ \beta_j) - u^2 v^2 \tag{16}$$

$$\times (a_i^+ a_j^+ \beta_j^+ \beta_i^+ - \beta_i \beta_j a_j a_i - a_i^+ a_j^+ a_j a_i - \beta_i^+ \beta_j^+ \beta_j \beta_i$$

$$-\beta_i^+ a_j^+ a_j \beta_i - a_i^+ \beta_j^+ \beta_j a_i) + u^4 a_i^+ a_j \beta_j^+ \beta_i + v^4 \beta_i^+ \beta_j a_j^+ a_i] + H'.$$

In H' are gathered all the terms which contain three creation operators and one annihilation operator, or vice versa.

The ground state $|0\rangle$ of the system (vacuum) is determined by the condition

$$\alpha_j |0\rangle = \beta_j |0\rangle = 0. \tag{17}$$

The energy of the ground state of the system is obtained by using (16) and (17):

$$E_0 = \langle 0|H|0\rangle = \frac{\omega_0}{2}(v^2 - u^2)N - \frac{1}{2}u^2 v^2 \sum_{ij} B(\mathbf{r}_i - \mathbf{r}_j). \tag{18}$$

The coefficients u and v are determined from the relation

$$\left\langle 0 \left| \sum_j (a_{j+1/2}^+ a_{j+1/2} - a_{j-1/2}^+ a_{j-1/2}) \right| 0 \right\rangle = N(v^2 - u^2) = n_+ - n_-, \tag{19}$$

using the condition $v^2 + u^2 = 1$, where $n_+$ and $n_-$ are the numbers of excited and unexcited nuclei in the crystal, respectively. As a result we have

$$v^2 = n_+ / N, \quad u^2 = n_- / N. \tag{20}$$

The result (20) is perfectly obvious, since $v^2$ corresponds to the probability of the presence of excitation in the given site, and $u^2$ yields the probability of absence of excitation.

We represent the ground state energy (18) in the form

$$E_0 = E_0^{(1)} - i\frac{\Gamma_R}{2},$$

$$E_0^{(1)} = \frac{\omega_0}{2}(v^2 - u^2)N - \frac{1}{2}u^2v^2 \text{Re}\left(\sum_{ij} B(\mathbf{r}_i - \mathbf{r}_j)\right),$$

$$\Gamma_R = \text{Im}\left(\sum_{ij} B(\mathbf{r}_i - \mathbf{r}_j)\right) = 2\pi \frac{n_+ n_-}{N^2}\sum_\lambda \rho_\lambda(\omega_0)\int d\Omega_\mathbf{K} |H_{\mathbf{K}\lambda}|^2 \left|\sum_j e^{i\mathbf{K}\mathbf{r}_j}\right|^2,$$

where $\Gamma_R$ has the meaning of the radiation width for the entire crystal. If the nuclei are not polarized, then

$$\Gamma_R = \frac{\gamma_R}{4\pi} n_+ n_- e^{-W} \frac{1}{N^2}\int d\Omega_\mathbf{K} \left|\sum_j e^{i\mathbf{K}\mathbf{r}_j}\right|^2. \tag{21}$$

It is also obvious that if condition (5) is exactly satisfied the probability of emission in the direction of **K** is

$$dw/\Omega_\mathbf{K} = \frac{\gamma_R}{4\pi} e^{-W} n_+ n_-. \tag{22}$$

Formulas (21) and (22) agree with the results of the preceding section. Indeed, when $n_+ \ll n_-$, the radiation probability of the entire crystal can be obtained by simple multiplication of (10) by $n_+$. In the same manner as in the preceding section, we verify that

$$\Gamma_R \approx \gamma_R e^{-W} n_+ n_- N^{-2/3}/4\pi. \tag{23}$$

The expressions obtained are symmetrical with respect to $n_+$ and $n_-$. The maximum radiation intensity should be observed when $n_+ \sim n_- \sim N$; in this case

$$\Gamma_R \approx \gamma_R e^{-W} N^{4/3}/4\pi. \tag{24}$$

In the other case, when $n_+$ or $n_-$ are small,

$$\Gamma_R \approx \gamma_R e^{-W} N^{1/3}/4\pi. \tag{25}$$

In the preceding analysis we disregard absorption of the $\gamma$ quanta (extinction). The effective number of unexcited nuclei, which can participate in the coherent radiation is $N_{eff} \sim (l_{ext}/a)^3$. We can estimate $l_{ext}$ roughly, using the formulas given [8]

$$n^2 - \chi^2 - 1 = \frac{\omega - \omega_0}{(\omega - \omega_0)^2 + (\Gamma/2)^2} 2\pi\gamma_R \left(\frac{\lambdabar}{a}\right)^3,$$

$$n\chi = \frac{\Gamma/2}{(\omega - \omega_0)^2 + (\Gamma/2)^2} \pi\gamma_R \left(\frac{\lambdabar}{a}\right)^3, \quad \Gamma = \gamma_R + \gamma_C, \tag{26}$$

where n is the refractive index, and $K$ the extinction coefficient, which is connected with $l_{ext}$ by the relation $l_{ext} = \lambdabar/k$.

We are interested in the case when $\lambdabar/a \ll 1$; then we have for $\omega \approx \omega_0$

$$\chi \approx 2\pi(\gamma_R/\Gamma)\left(\frac{\lambdabar}{a}\right)^3, \tag{27}$$

and consequently

$$N_{eff} \sim \left[(\Gamma/2\pi\gamma_R)(a/\lambdabar)^2\right]^3, \tag{28}$$

If $\Gamma/\gamma_R = 1 + \alpha$ ($\alpha$ -conversion coefficient $\sim 10-10^2$), $\lambdabar = 0.5 \times 10^{-9} cm$, and $a = 2 \times 10^{-8} cm$, then

$$N_{eff} \sim 10^{10} - 10^{13}. \tag{29}$$

Accordingly, in all the preceding formulas we must substitute $N_{eff}$ in place of N. In particular, the estimate (25) will take the form

$$\Gamma_R \approx \gamma_R e^{-W} N_{eff}^{1/3}/4\pi. \tag{30}$$

Thus, for real values of the parameters the total width for the radiation by the entire crystal can become much greater than the width for the isolated nucleus (coherent broadening).

**CONCLUSION**

It is seen from the foregoing analysis that in the case when a nuclear exciton is produced the main characteristics of the spontaneous emission change noticeably: a) angular anisotropy appears, the y-quantum flux is concentrated in the b direction, and the probability of emission in this direction is $\sim N_{eff}^2$; b) the total width for radiatio~ by the entire crystal becomes longer than the width of the isolated nucleus by a factor $\sim e^{-W} N_{eff}^{1/2}$.

Thus, if the conditions for the realization of a nuclear exciton are satisfied then: 1) sharply directed beams of monochromatic $\gamma$ quanta can be obtained; 2) the lifetimes of the nuclear isomers can be appreciably reduced when these isomers are placed in a crystal consisting of unexcited nuclei of the same sort.

To check on the possibility of existence of a nuclear exciton, it is necessary to vary the lattice constant in any one of the crystallographic directions until the condition $\mathbf{K} = 2\pi\mathbf{b}$ is satisfied. The nuclear exciton can be observed from the resultant angular anisotropy of the y quanta, or from the broadening of the Mossbauer emission line. The lattice constant can be changed by applying pressure or by varying the temperature.